\documentclass[onecolumn,secnumarabic,amssymb, nobibnotes, aps, prb]{revtex4-1}

\usepackage{graphicx}
\usepackage{subfigure}
\begin{document}
\title{Electronic transport calculations for lightly-doped thermoelectrics using density functional theory: Application to high-performing Cu-doped zinc antimonides}%

\author{Alireza Faghaninia}%
\affiliation{Energy, Environmental, and Chemical Engineering, Washington University in St. Louis}
\author{Cynthia S. Lo}%
\email[]{clo@wustl.edu}
\affiliation{Energy, Environmental, and Chemical Engineering, Washington University in St. Louis}
\date{\today}%

\begin{abstract}
We propose a new method for accurately calculating electrical transport properties of a lightly-doped thermoelectric material from density functional theory (DFT) calculations, based on experimental data and density functional theory results for the corresponding undoped material.  We employ this approach because hybrid DFT calculations are prohibitive for the large supercells required to model low dopant concentrations comparable to those achieved experimentally for high-performing thermoelectrics.  Using zinc antimonide as our base material, we find that the electrical transport properties calculated with DFT and Boltzmann transport theory exhibit the same trends with changes in chemical potential as those computed with hybrid DFT, and propose a fitting algorithm that involves adjusting the computed Fermi energy so that the resulting Seebeck coefficient trends with temperature match experimental trends.  We confirm the validity of this approach in reproducing the experimental trends in electrical conductivity and Seebeck coefficient versus temperature for Bi-doped $\beta-$Zn$_4$Sb$_3$.  We then screen various transition metal cation dopants, including copper and nickel, and find that a Cu dopant concentration of 2.56\% in Zn$_{39}$Sb$_{30}$ exhibited a 14\% increase in the thermoelectric power factor for temperatures between 300-400 K.  We thus propose that transition metal dopants may significantly improve the thermoelectric performance of the host material, compared to heavy and/or rare-earth dopants.
\end{abstract}
\maketitle

\section{Introduction}
\label{introduction}

Doping of thermoelectric materials with transition metals is expected to significantly enhance their electrical transport properties, but validation of this hypothesis and screening of new materials require advances in computational power and/or improvements in theoretical prediction of these properties.  The efficiencies of these thermoelectric materials, which can be used in a "solid-state heat engine" \cite{41} to convert temperature gradients into voltage differences, are quantified by a dimensionless figure of merit, defined as:
\begin{eqnarray*}
ZT & = & \frac{S^2 \sigma}{k} T
\end{eqnarray*}
where $\sigma$ is the electrical conductivity, $S$ is the Seebeck coefficient, or thermopower, $T$ is the temperature, and $k$ is the thermal conductivity of the material.  The aim is to improve the thermoelectric power factor, or numerator of $Z$, enough so that $ZT > 1$ and these materials could conceivably be used for large-scale power generation and solar energy capture and conversion. 

$\beta$-Zn$_4$Sb$_3$ is one of the better-performing thermoelectric materials in the 450-800 K (i.e., medium) temperature range \cite{68,71,74,66,handbook1} that is suitable for utilizing waste heat from industrial power plants and concentrated solar thermal energy.  Its crystal structure is quite complex, with numerous defects and vacancies that create short-range and long-range disorder \cite{65}, which accounts for its extremely low thermal conductivity and high thermoelectric figure of merit \cite{65,64}.  It also is thought to exhibit similar electronic structure to ZnSb, and an analysis of both structures will be presented in this paper.

While numerous computational studies have been performed on ZnSb \cite{58,67,69,74,75,78,89}, none of them appear to correctly obtain the experimental electronic band gap of the bulk material.  The calculated indirect band gap has been reported to range from $E_g = 0.05-0.3 \ \text{eV}$, while the experimental $E_g = 0.61 \ \text{eV}$ at $T = 4.2 \ \text{K}$ \cite{75,73,94,95,97} and $E_g = 0.5 \ \text{eV}$ at $T = 300 \ \text{K}$ \cite{94,95}; this result is a consequence of both the known systematic underestimation of the band gap, due to the discontinuity of the electron exchange potential in the Kohn-Sham density funcitonal theory (DFT) formulation \cite{101,105}, and electron delocalization caused by the DFT self-interaction error \cite{hse1,100,102,103}.  Using a fraction of the Hartree-Fock exchange potential in the hybrid DFT formulation \cite{hse1} results in an increase in the computed band gap that approaches the experimental values, but its use is computationally prohibitive for systems with large numbers of atoms.

Doped zinc antimonides are a prime example of the type of system that is large in its required unit cell size, but potentially quite valuable as candidates for high-performing thermoelectrics.  Both $n$-type dopants (e.g., Li \cite{78,126,86}, Na \cite{86}, Zr \cite{86}, Hf \cite{86}, Sn \cite{86,114,119}, Te \cite{69,96,118,119}, and Tl \cite{86}) and $p$-type dopants (e.g., Cr \cite{121}, Mn \cite{121}, Cu \cite{120}, Ag, Au, Pb, Ga \cite{119}, In \cite{112,119}, Bi \cite{60,61}, Eu \cite{124}, Ho \cite{113}, Yb \cite{124} , and Te) have been successfully incorporated into ZnSb -- raising hopes for a low-strain $p-n$ junction in a thermoelectric generator.  Complex alloys, such as Ru$_x$Sb$_y$Zn$_z$ \cite{115}, have been reported to exhibit both $n$-type and $p$-type behavior, so the exact electronic structure remains to be confirmed.  Moreover, AZn$_2$Sb$_2$ Zintl alloys have been shown to exhibit $p$-type (e.g., A = Ca, Sr, Yb, Eu) behavior \cite{124}. In all of these Zintl compounds, the hole carriers are formed when the A atoms donate electrons -- leaving holes behind. We thus hypothesize that improved $p$-type doping of ZnSb and Zn$_4$Sb$_3$ -- resulting in higher carrier concentrations and higher electrical conductivity -- is achieved by using substitutional metal dopants that are more electropositive than Zn; thus, candidate dopants include Fe, Co, Ni, and Cu.  We presume that if the host material is only lightly doped, then any changes in carrier scattering would be slight enough to not affect the thermal conductivity, while still providing observable increases to the electrical transport properties.

In this paper, we present a fitting algorithm that predicts trends in electrical transport properties of thermoelectric materials upon $p$-doping, and apply it to the study of undoped ZnSb, and $p$-type Ni- and Cu-doped Zn$_4$Sb$_3$.  We note that potential "hole-killers" in $p$-doped zinc antimonides include Zn interstitials and Sb vacancies; these defects may lower the hole carrier concentration \cite{ZnS11}.  Thus, we also calculate the energy of formation for each of these defects.  We propose that this approach may be used to screen new materials for superior thermoelectric properties, and may also be applied to the study of other electrically-conductive materials.

\section{Model and Methodology}
\label{theory}
\subsection{Band structure calculation of undoped ZnSb}
\label{ZnSbsubsection}

We begin by computing the electronic band structure of undoped ZnSb, using Kohn-Sham DFT (KS-DFT), hybrid DFT, and $GW$ methods to later establish a fitting algorithm that may be used to predict the electrical transport properties of related compounds, such as Zn$_4$Sb$_3$.  The initial unit cell coordinates are consistent with the orthorhombic Pb$ca-D_{2h}$ space group, with $\left| \mathbf{a} \right| = 6.2016 \ \text{\AA}$, $\left| \mathbf{b} \right| = 7.7416 \ \text{\AA}$, and $\left| \mathbf{c} \right| = 8.0995 \ \text{\AA}$.  The structure was optimized using density functional theory (DFT) \cite{hohenberg_physrev_1964,kohn_physrev_1965}, as implemented in the Vienna Ab Initio Simulation Package (VASP) \cite{vasp1,vasp2,vasp3,vasp4}.  The generalized gradient approximation of Perdew, Burke, and Ernzerhof \cite{gga1,gga2} was used to express the exchange-correlation potential, and the Projector Augmented Wave (PAW) potentials \cite{paw1,paw2} are used to represent the all-electron valence wavefunctions as accurately as possible with reasonable computational cost.  The first Brillouin zone was sampled with a $6 \times 5 \times 5$ $k$-point mesh, which corresponds to 150 $k$-points.  The kinetic energy cutoff for the plane wave basis set was set to 400 eV after extensive trials.  The band structure was computed along seven high-symmetry $k$-points in the first Brillouin zone, with 40 $k$-points between each pair of high-symmetry points.  All energies were converged to $10^{-5} \ \text{eV}$ and all forces were converged to $0.02 \ \text{eV}/\text{\AA}$.  Upon DFT optimization, the changes to the lattice dimensions of the unit cell are: $\Delta \left| \mathbf{a} \right| = +1.15\%$, $\Delta \left| \mathbf{b} \right| = +0.9\%$, and $\Delta \left| \mathbf{c} \right| = +1.41\%$.  Upon HSE optimization, the changes to the lattice dimensions of the unit cell are: $\Delta \left| \mathbf{a} \right| = +1.37\%$, $\Delta \left| \mathbf{b} \right| = -0.16\%$, and $\Delta \left| \mathbf{c} \right| = +1.46\%$

The $GW$ calculations are computationally intensive, so a smaller $k$-point mesh of $3 \times 3 \times 3$ was used to sample the first Brillouin zone, and the same line mesh used in the DFT and HSE calculations is used to compute the band structure.  The $GW$ band structure calculations are performed using the maximally-localized Wannier functions (MLWFs) interpolation, as implemented in VASP and the Wannier code \cite{93}.

The electrical transport properties, including electrical conductivity and Seebeck coefficient, were computed using the BoltzTraP code \cite{50}, which is based on Boltzmann semi-classical theory and Fourier interpolation of the calculated bands.  The band energies were differentiated to find the group velocities of the electrons.  One shortcoming of this method is that the output file from a VASP calculation lists the energy states at each $k$-point without regard to which band they belong; when two bands cross, they cannot be discerned with confidence, and this may lead to errors in band assignment.  This error may be minimized by increasing the density of the $k$-point mesh.  Thus, we used a $k$-point mesh of $28 \times 28 \times 28$ (i.e., 3375 $k$-points in the irreducible Brillouin zone, or IBZ) for the non-self-consistent DFT calculations, and a $k$-point mesh of $12 \times 12 \times 12$ (i.e., 343 $k$-points in the IBZ) for the non-self-consistent HSE calculations, when calculating the properties; we note that these meshes are much denser than those required to merely plot the band structure, and appear to be sufficient to obtain the desired properties with reasonable computational cost.

\subsection{Band structure calculation of $\beta-$Zn$_4$Sb$_3$}
\label{Zn4Sb3subsection}
\begin{figure}
\begin{center}
            \includegraphics[width=0.5\textwidth]{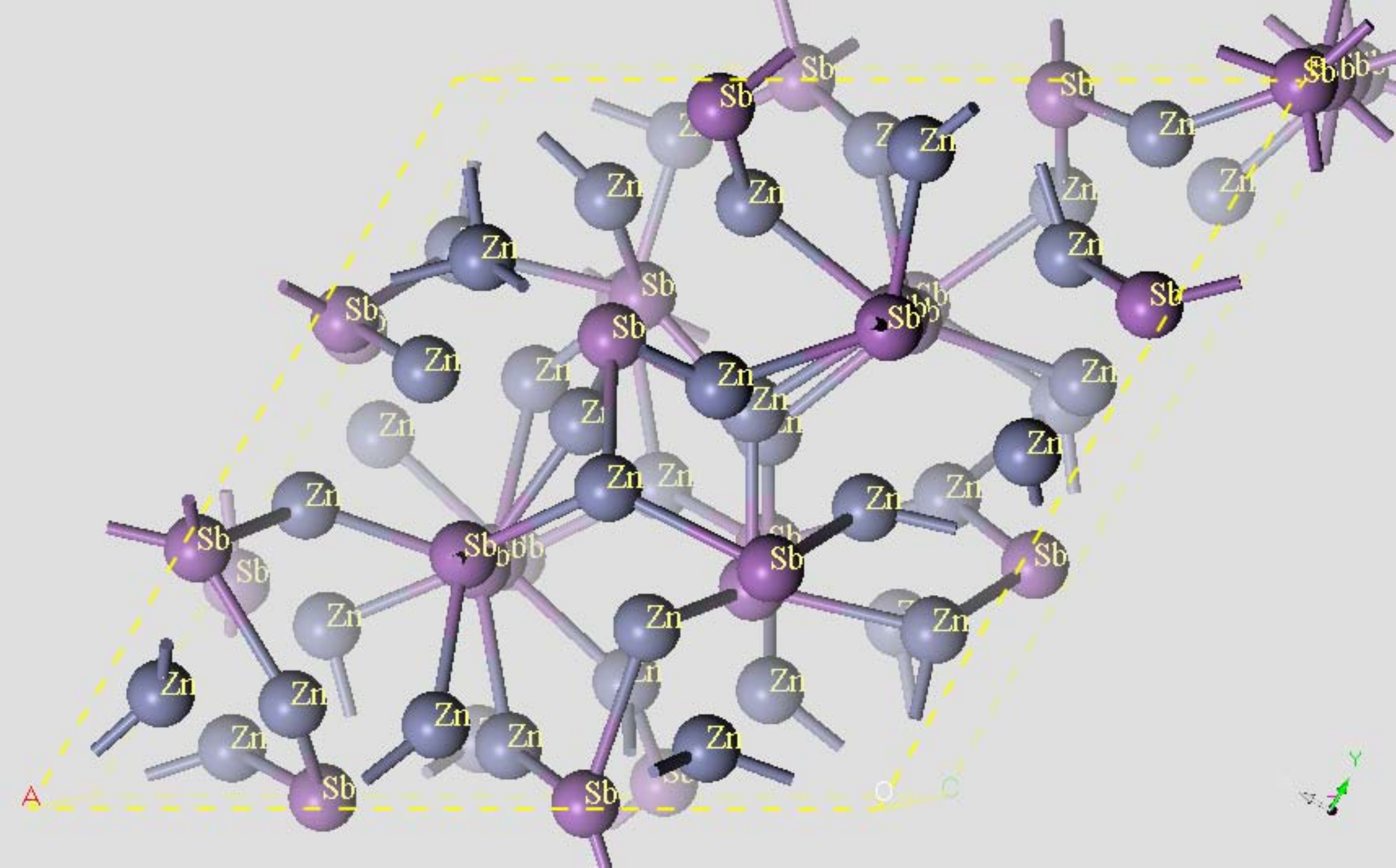}
   \end{center}
   \caption{ \label{fig:unitcellZn4Sb3}Geometry-optimized structure of Zn$_{39}$Sb$_{30}$ unit cell.}
\end{figure}

The crystal structure of $\beta$-Zn$_4$Sb$_3$ is consistent with the disordered rhombohedral $R\bar{3}c$ space group\cite{63}, and contains partially-occupied sites, which are not possible to explicitly simulate using DFT.  Instead, we constructed the structure by forming a large supercell and selectively removing and adding Zn atoms according to published procedures \cite{88} to form Zn$_{39}$Sb$_{30}$ (Figure \ref{fig:unitcellZn4Sb3}).  In particular, the main Zn1 site (36f) is only partially occupied, with Zn-A having 0.898 occupancy, Zn-B having 0.067 occupancy, Zn-C having 0.056 occupancy, and Zn-D having 0.026 occupancy; this results in an overall stoichiometry of Zn$_{3.77}$Sb$_{3}$.

The geometry optimization was performed according to the same procedure as in Section \ref{ZnSbsubsection}, but the energy cutoff was chosen to be 359.74 eV, since no major changes were observed upon increasing it to 400 eV; the lower energy cutoff facilitates calculations on the larger supercell without loss of accuracy.  While no symmetry is observed in the optimized $\beta$-Zn$_4$Sb$_3$ unit cell (i.e., it is amorphous), we select the same high-symmetry $k$-points from the $R\bar{3}c$ symmetry group reported with partial occupancies \cite{63}.  Self-consistent calculations were performed using a $k$-point mesh of $4 \times 4 \times 4$, which we deem sufficiently large since the use of a larger $6 \times 6 \times 6$ $k$-point mesh changed the total energy of the unit cell by only 0.001 eV.  Density of states (DOS) calculations were performed using this $6 \times 6 \times 6$ $k$-point mesh, and non-self-consistent calculations, as needed to evaluate the Fermi energy and calculate electrical transport properties, were performed using an $8 \times 8 \times 8$ $k$-point mesh.  These mesh sizes represented the smallest number of $k$-points that would still provide converged property results; furthermore, the unit cell size is sufficiently large that these $k$-point meshes are sufficient to achieve convergence.

\subsection{Band structure and property calculation of 2.56\% Ni-, Cu-, and Bi-doped Zn$_4$Sb$_3$}
\label{MZn4Sb3subsection}

We propose the following methodology to compute the electrical transport properties of a lightly-doped compound, as modeled by a large supercell (e.g., M$_x$Zn$_{38}$Sb$_{30}$, where M is the substitutional dopant):
\begin{enumerate}
\item{Perform self-consistent and non-self-consistent calculations on undoped supercell to obtain all energy eigenstates.}
\item{\label{fermi_real}Calculate the average highest occupied band energy (i.e., Fermi energy); the weighted average should be based on the weight of each $k$-point.}
\item{\label{fermi_scaled}Manually adjust the Fermi energy in the BoltzTraP calculation to fit the computed Seebeck coefficient versus temperature to available experimental data for the undoped compound.}
\item{Use the adjusted Fermi energy to calculate $\frac{\sigma}{\tau}$, where $\tau$ is the electronic relaxation time, using the BoltzTraP code.  Calculate $\tau$ by fitting the computed electrical conductivity versus temperature to experimental data for the undoped compound.}
\item{Create a doped supercell by substituting one Sb atom for Bi, or one Zn atom for Cu or Ni.  Structurally optimize the unit cell with DFT.}
\item{Use the difference in Fermi energies between steps \ref{fermi_real} and \ref{fermi_scaled} to calculate the new chemical potential, $\mu$, for the doped supercell, following the procedure in step \ref{fermi_real}.}
\item{Calculate the Seebeck coefficient and electrical conductivity for the doped supercell, using the computed $\mu$.}
\end{enumerate}
The Fermi level calculated in step \ref{fermi_scaled} represents the Fermi level of the "real" material with defects and impurities. In other words, we are calibrating our calculated properties to a specific experiment by changing the Fermi level. For a p-type material, this Fermi level is close to the valence band as described in step \ref{fermi_real}; decreasing the Fermi energy, and hence moving the Fermi level deeper inside the valence band results in more p-type behavior (i.e. more acceptor impurities), while increasing the Fermi energy results in more n-type behavior (i.e. more donor impurities). For naturally n-type materials, the calibration of step \ref{fermi_scaled} should be done for n-type conductivity; also, the lowest unoccupied band energy should be used. Our motivation for developing this approach is to rapidly screen potential dopants that may enhance electrical transport properties of the host material.  

\section{Results and discussion}
\label{results}
\begin{figure}
\begin{center}
            \includegraphics[width=0.45\textwidth]{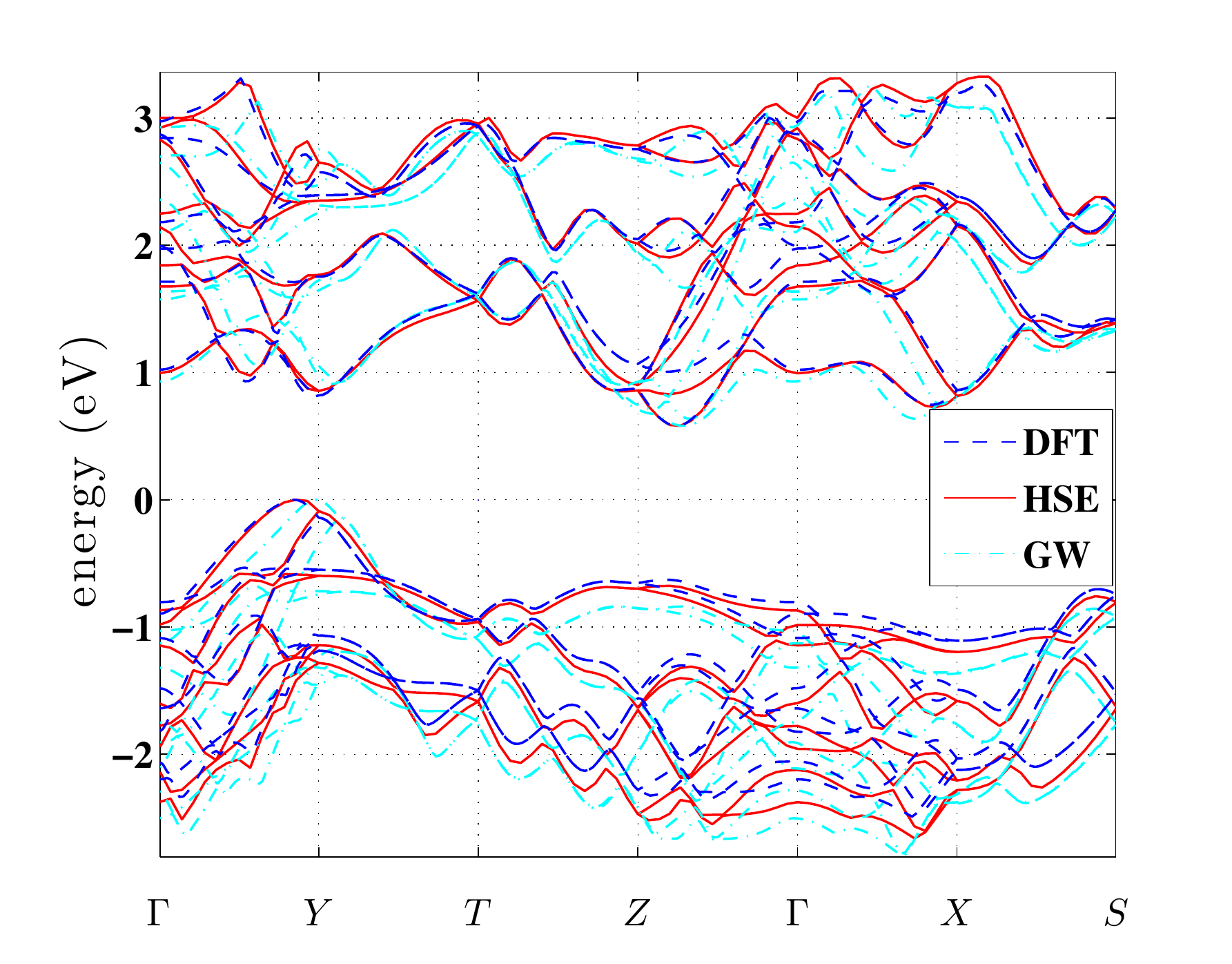}
   \end{center}
   \caption{\label{fig:DFT_HSE_GW}The difference between band energies of ZnSb calculated with DFT, HSE, and $GW$, after fitting the VBM and band gap from DFT and $GW$ to that from HSE}
 
\end{figure}

\begin{figure}
        \subfigure[]{
            \label{fig:DFTonHSE_s}
            \includegraphics[width=0.45\textwidth]{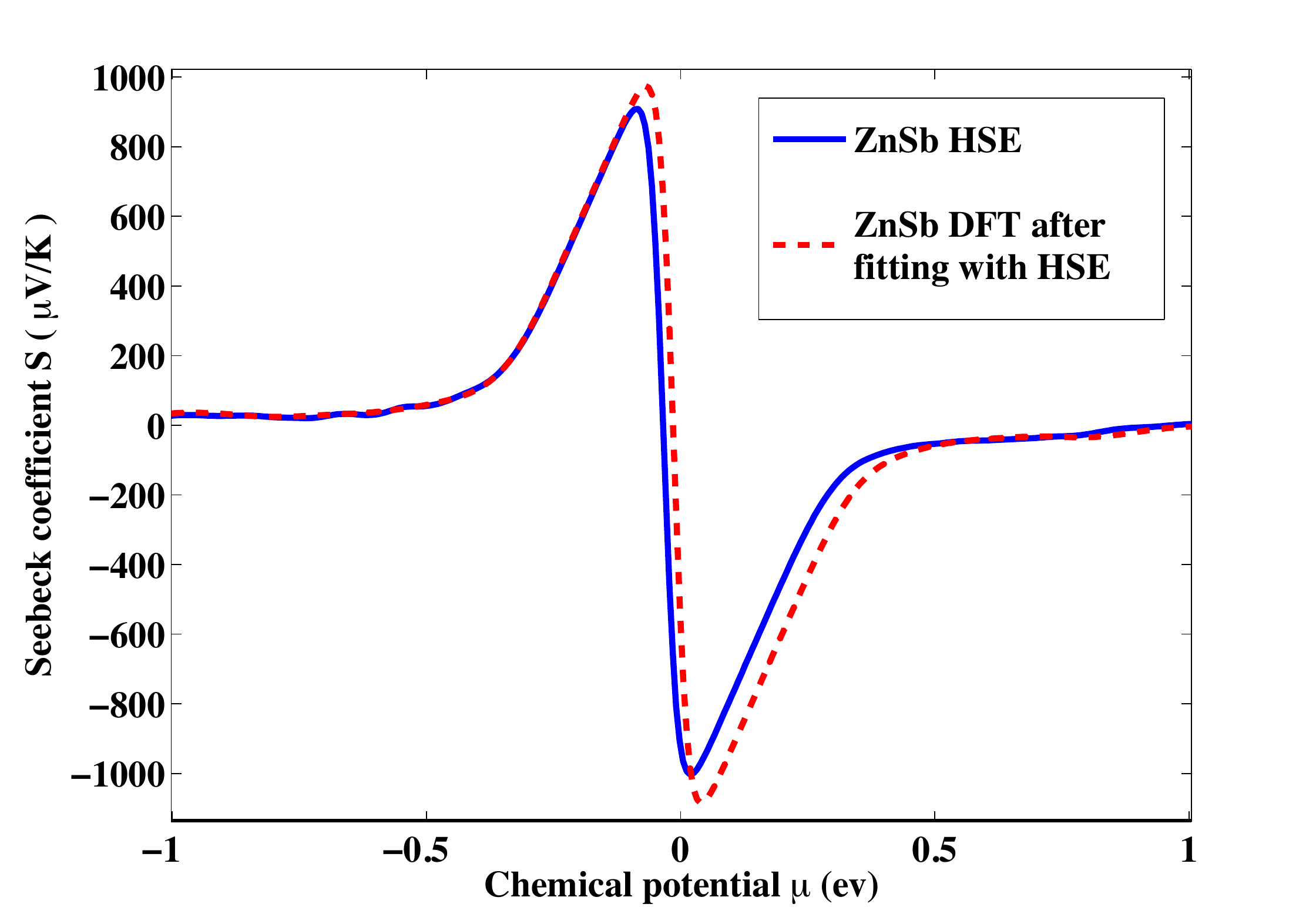}}
       \\ 
        \subfigure[]{
            \label{fig:DFTonHSE_e}
            \includegraphics[width=0.45\textwidth]{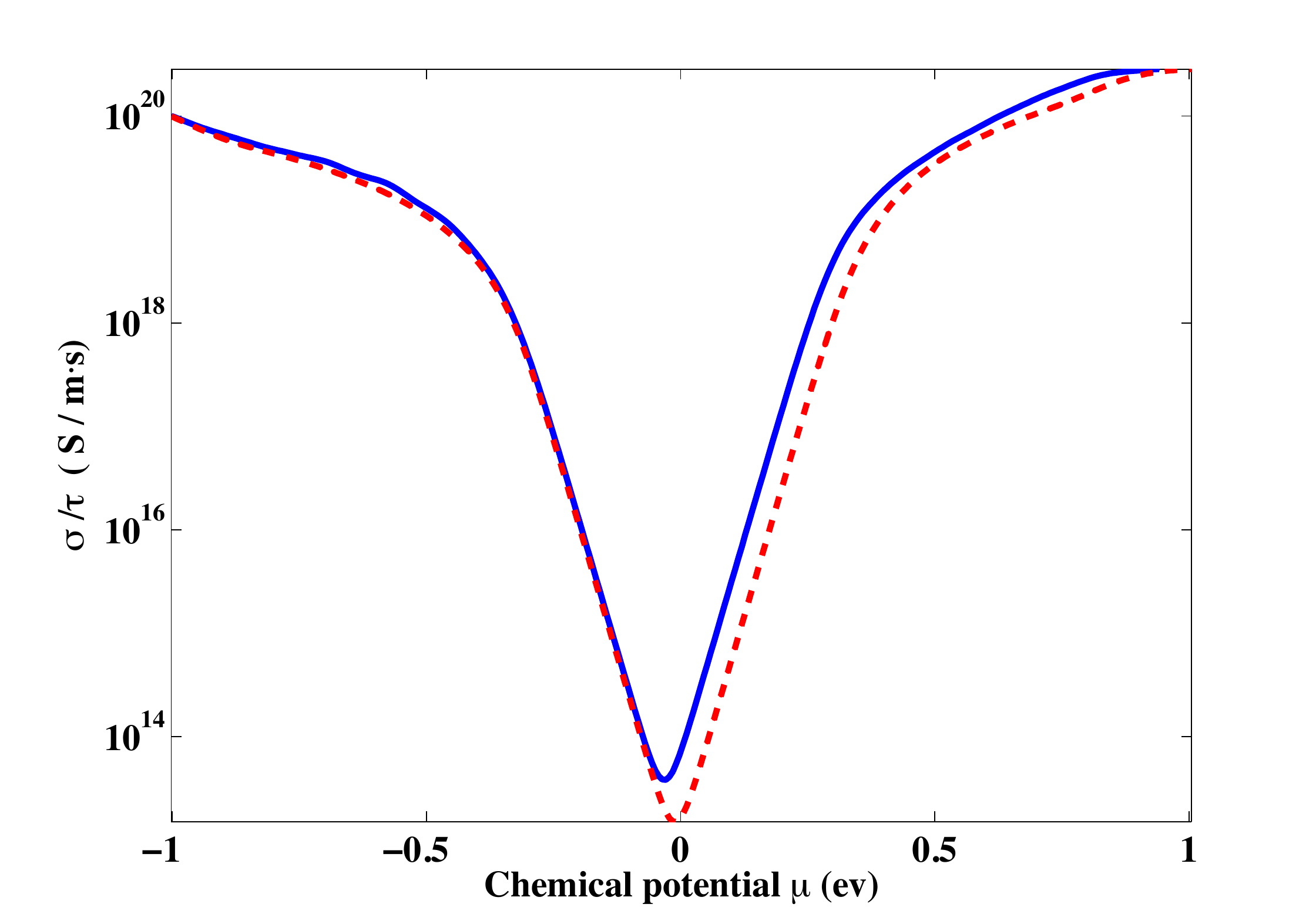}}
    \caption{\label{fig:DFTonHSE}Seebeck coefficient \subref{fig:DFTonHSE_s} and Electrical conductivity \subref{fig:DFTonHSE_e} divided by relaxation time calculated at 300 K, comparing the hybrid HSE functional to Kohn-Sham DFT with the VBM maximum and band gap fitted to those calculated by HSE. The Fermi energy has been set to the midgap calculated by HSE in both cases.}
\end{figure}

Since the Kohn-Sham DFT formulation contains a discontinuity in the exchange potential, this results in a systematic underestimation of the electronic band gap.  The magnitude of the underestimation is proportional to this discontinuity, and disappears in a band structure calculation for a metal \cite{101, 104}.  As an example, the DFT-calculated band gap for ZnSb is $E_g = 0.04 \ \text{eV}$, while those calculated using the $GW$ method and the HSE hybrid functional are $E_g = 0.39 \ \text{eV}$ and $E_g = 0.58 \ \text{eV}$, respectively, which are much closer to the experimental value of $E_g = 0.61 \ \text{eV}$ \cite{77}. The shifting of the Fermi level and associated valence bands results in the same rate of change in the calculated DFT properties as would be observed using the HSE band structure.  We note that the use of the HSE potential affects the band shapes and curvatures, especially at the valence and conduction band edges, so that the eigenstates are not simply shifted relative to the DFT values (Figure \ref{fig:DFT_HSE_GW}).  We too have utilized the scissor operator to shift the DFT bands in such a way that the indirect band gap and valence band maximum (VBM) are fit to the corresponding HSE band gap and VBM, and the Fermi level is set to the midgap of the HSE eigenstates (Figure \ref{fig:DFTonHSE}).  Although the values of the DFT+scissor and HSE properties differ in some regions, their rate of change is the same in all regions.  Therefore, we are confident that by fitting DFT bands to the experimental data for Seebeck coefficient and electrical conductivity, we can accurately reproduce trends in these properties upon doping of the host material without resorting to expensive HSE or $GW$ calculations.

\subsection{Validation of proposed fitting algorithm}
\begin{figure}
\begin{center}
            \includegraphics[width=0.45\textwidth]{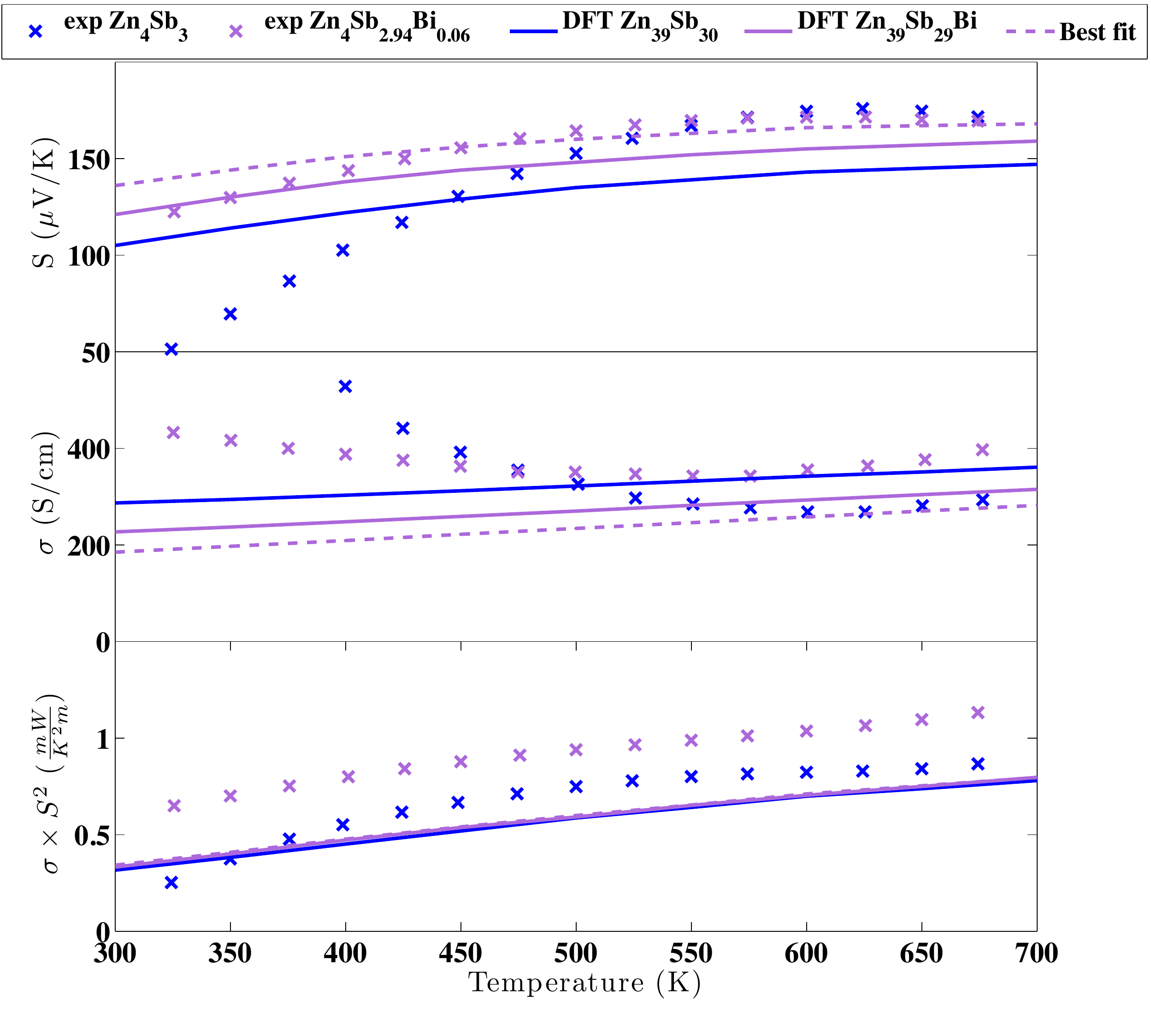}
   \end{center}
   \caption{\label{fig:Biresults}2.56\% Bi doped Zn$_4$Sb$_3$ calculated and experimental \cite{60} thermoelectric properties: Seebeck coefficient, electrical conductivity, and power factor respectively. The best fit that could be achieved with BoltzTraP \cite{50} is also shown.}
\end{figure}

To validate the proposed fitting algorithm, we first performed property calculations on Zn$_{39}$BiSb$_{29}$, as this compound has been previously synthesized and characterized.  We evaluated all possible Sb substitutional sites for the Bi atom, the average total energy of these 30 configurations was $-165.82\pm0.005 \ \text{eV}$. We chose the configuration that results in the lowest total energy of $-166.02 \ \text{eV}$.  The energy of formation for the Bi substitutional defect is 0.43 eV.  The computed properties, as fit to the published experimental data \cite{60}, are shown in Figure \ref{fig:Biresults}. In this Figure, in addition to properties calculated via the method described in \ref{MZn4Sb3subsection}, the best fit possible with BoltzTraP is also included.  This corresponds to directly fitting the Fermi level of the doped supercell to the experimental data for the Seebeck coefficient of the doped compound. 
Although the fitted Seebeck coefficient and electrical conductivity from BoltzTraP vary slightly from the experimental data, the calculated and fitted thermoelectric power factor are nearly identical.  We presume that the origin of the systematic error is due to the relaxation time, $\tau$, being unchanged between the undoped calculation and the doped calculation.  This constant $\tau$ assumption fails to account for property changes with temperature.  Furthermore, carrier scattering mechanisms are not explicitly included in the $\tau$ determination.  Finally, we note that the properties depends strongly on the synthesis method employed; for the case of Bi-doped Zn$_4$Sb$_3$, hot pressing of the material at different temperatures \cite{71} results in more linear trends in the properties, compared to vacuum melting followed by hot pressing.  Even with these shortcomings, the fitting algorithm accurately predicts changes in electrical transport properties upon doping of the host material.

\subsection{Property calculations for M$_x$Zn$_{38}$Sb$_{30}$}
\begin{figure*}
    \begin{center}
        \subfigure[Zn$_{39}$Sb$_{30}$, band gap: 0.257 eV]{
            \label{fig:bs_undoped}
            \includegraphics[width=0.3\textwidth]{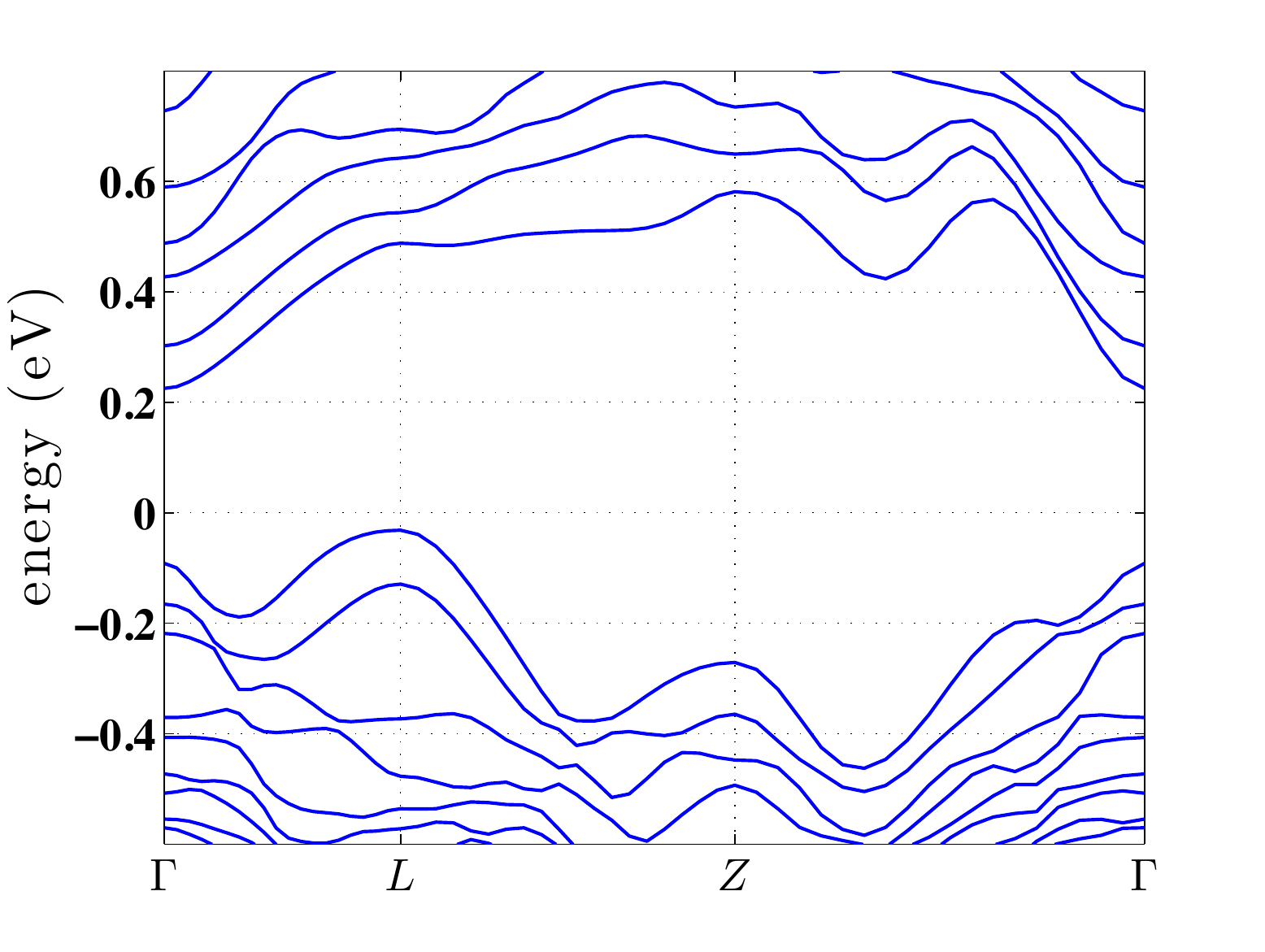}}
         \subfigure[NbZn$_{38}$Sb$_{30}$, band gap: 0.027 eV]{
           \label{fig:bs_Nb}
           \includegraphics[width=0.3\textwidth]{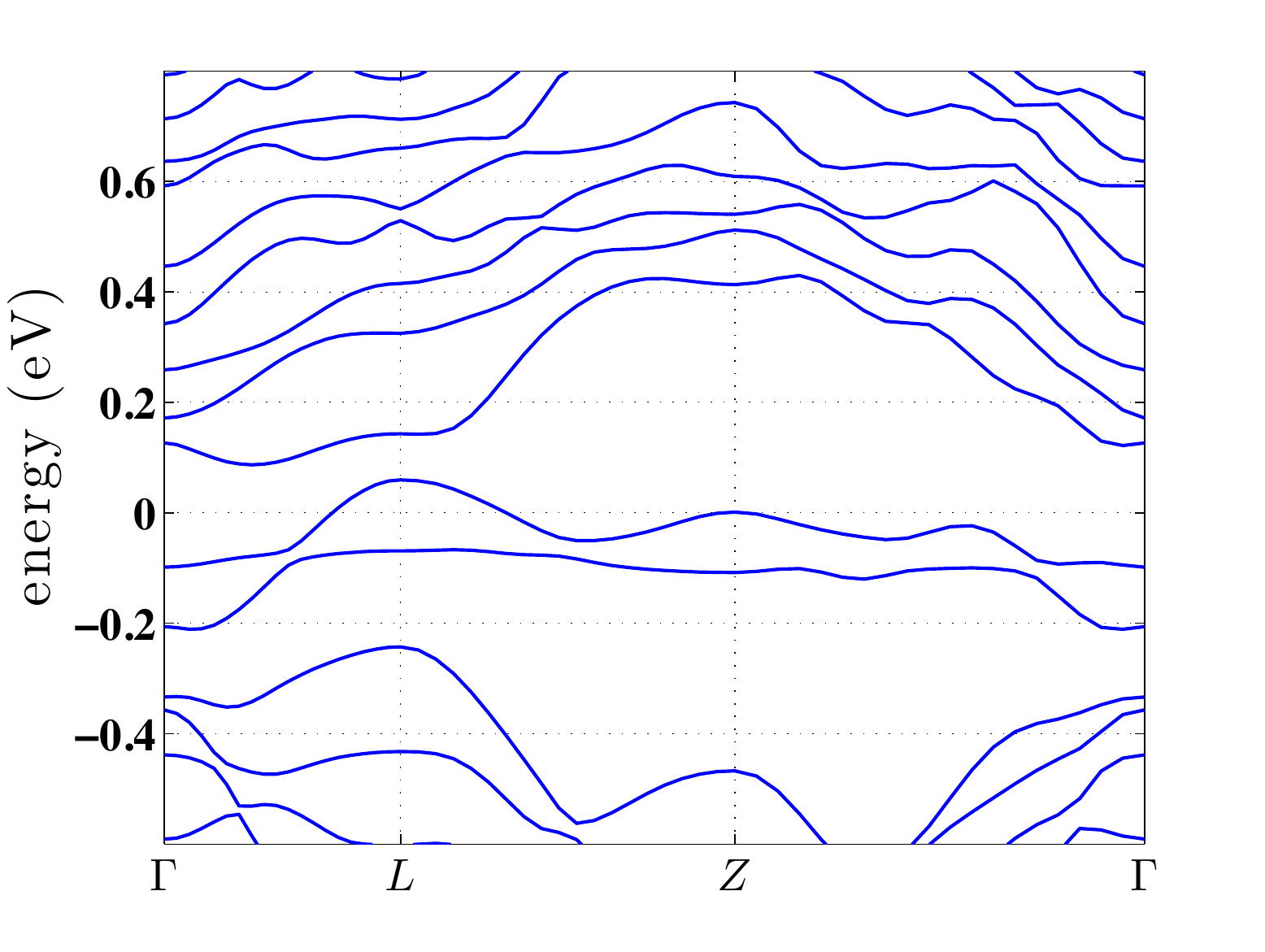}}
                    \subfigure[CuZn$_{38}$Sb$_{30}$, band gap: 0.282 eV]{
           \label{fig:bs_Cu}
           \includegraphics[width=0.3\textwidth]{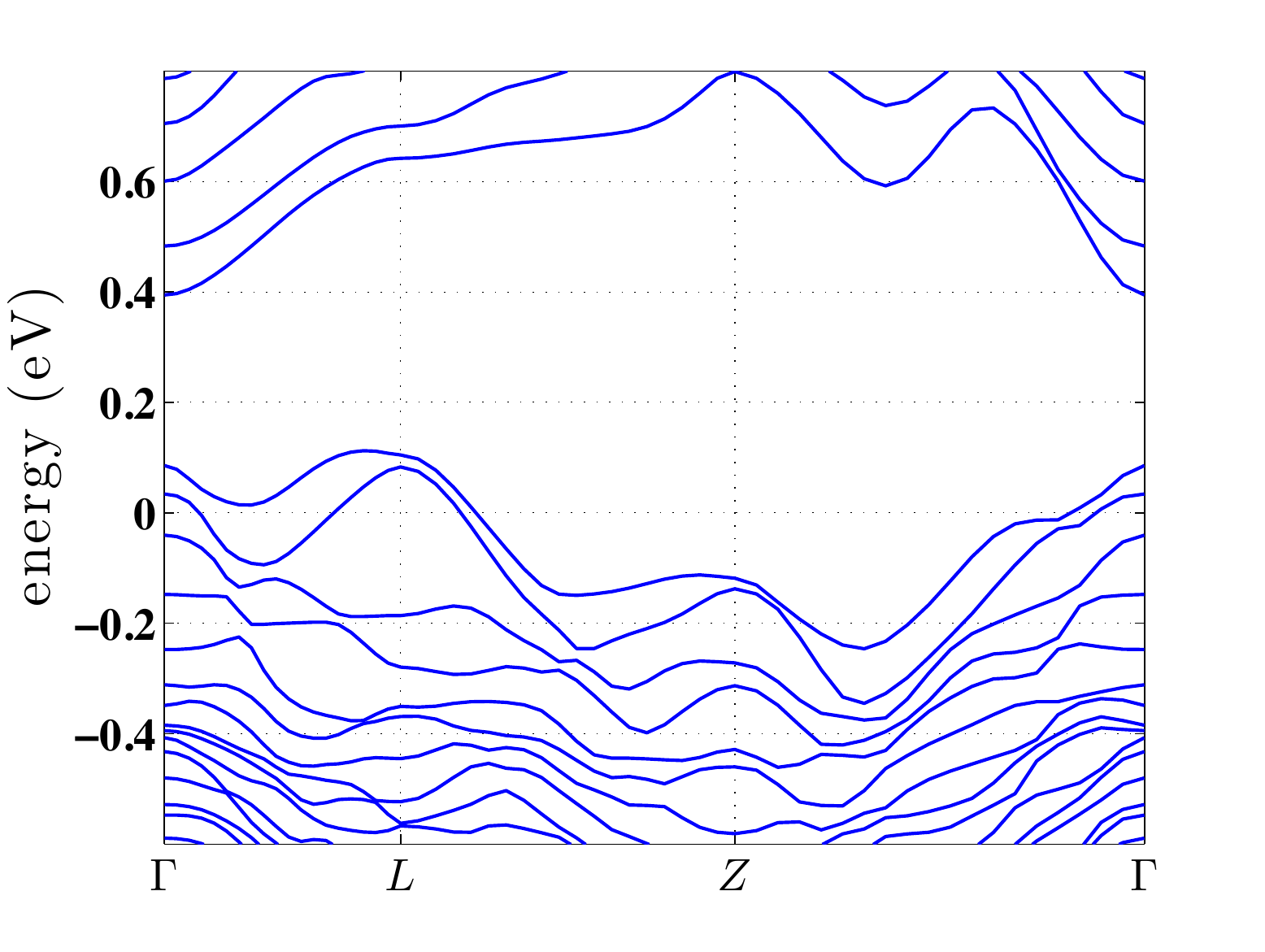}}
       \\ 
        \subfigure[FeZn$_{38}$Sb$_{30}$, band gap: 0.018 eV]{
            \label{fig:bs_Fe}
            \includegraphics[width=0.3\textwidth]{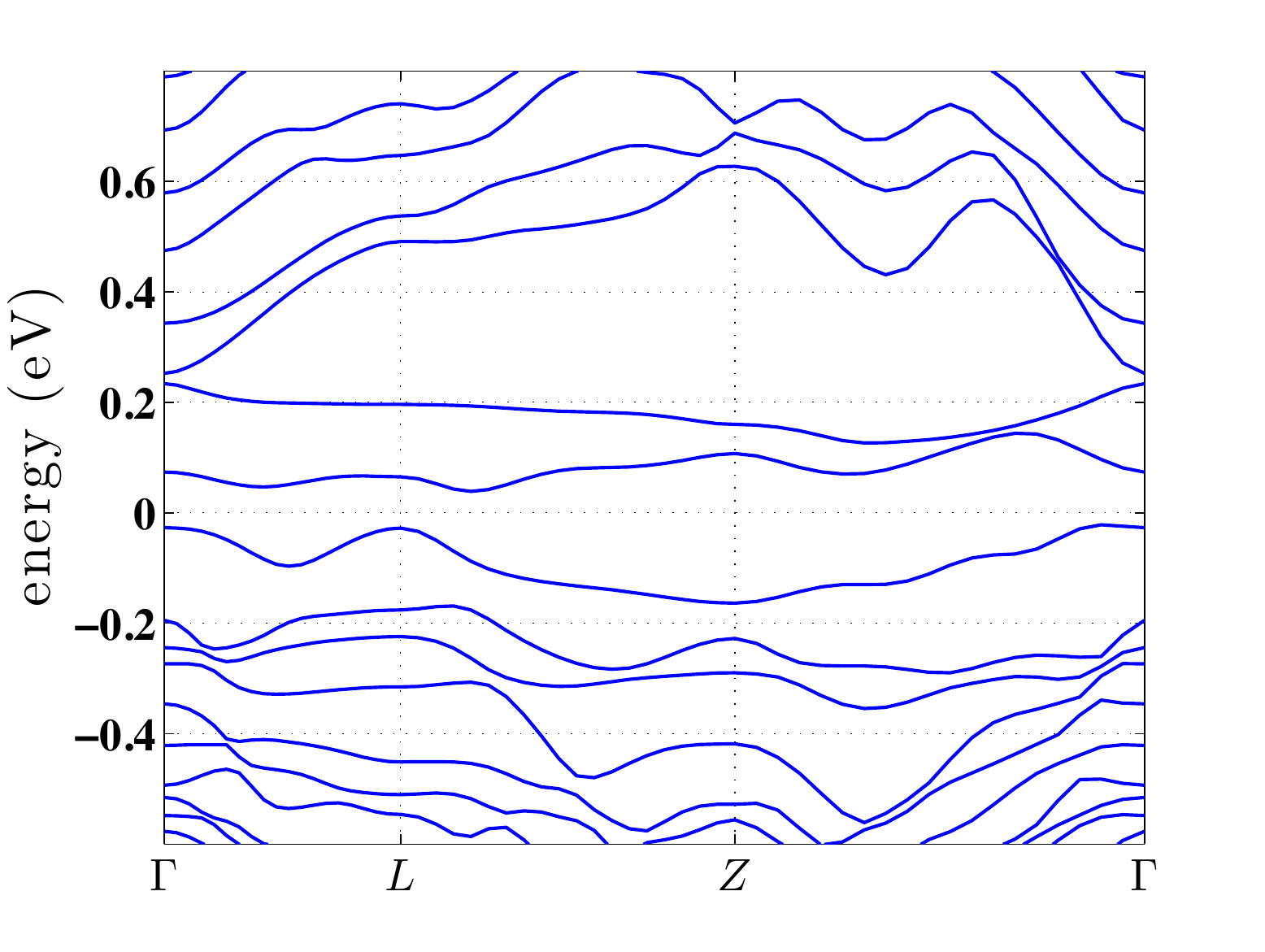}}
        \subfigure[CoZn$_{38}$Sb$_{30}$, band gap: 0.107 eV]{
            \label{fig:bs_Co}
           \includegraphics[width=0.3\textwidth]{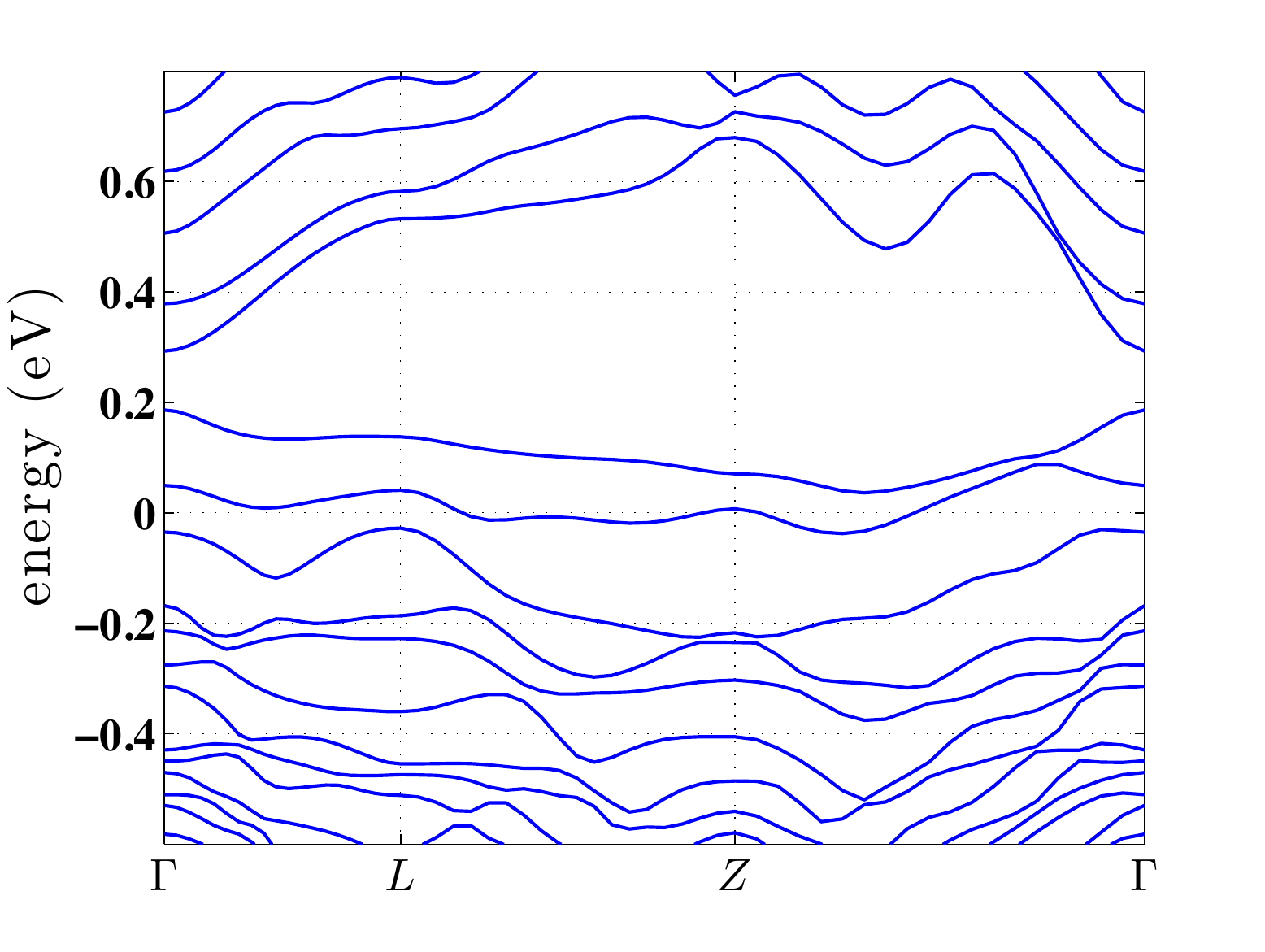}}
                     \subfigure[NiZn$_{38}$Sb$_{30}$, band gap: 0.2 eV]{
           \label{fig:bs_Ni}
           \includegraphics[width=0.3\textwidth]{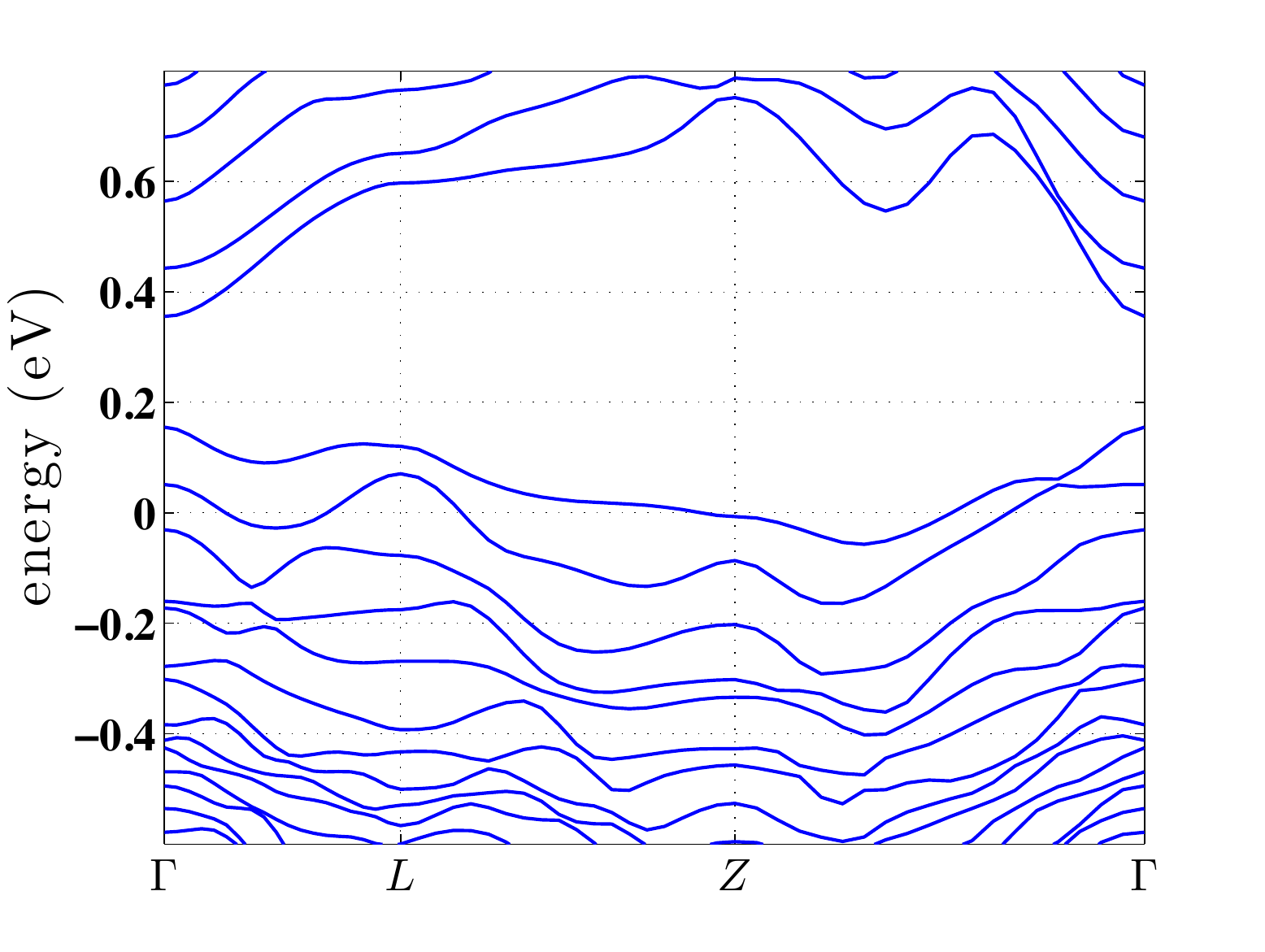}}
    \end{center}
    \caption{\label{fig:bs1}DFT calculated band structure and band gap of Zn$_{39}$Sb$_{30}$ and 2.56\% M-doped in the main site of MZn$_{38}$Sb$_{30}$; M is Nb, Fe, Co, Ni, or Cu, with R-3 symmetry}
\end{figure*}

\begin{figure}
\begin{center}
          
            \includegraphics[width=0.45\textwidth]{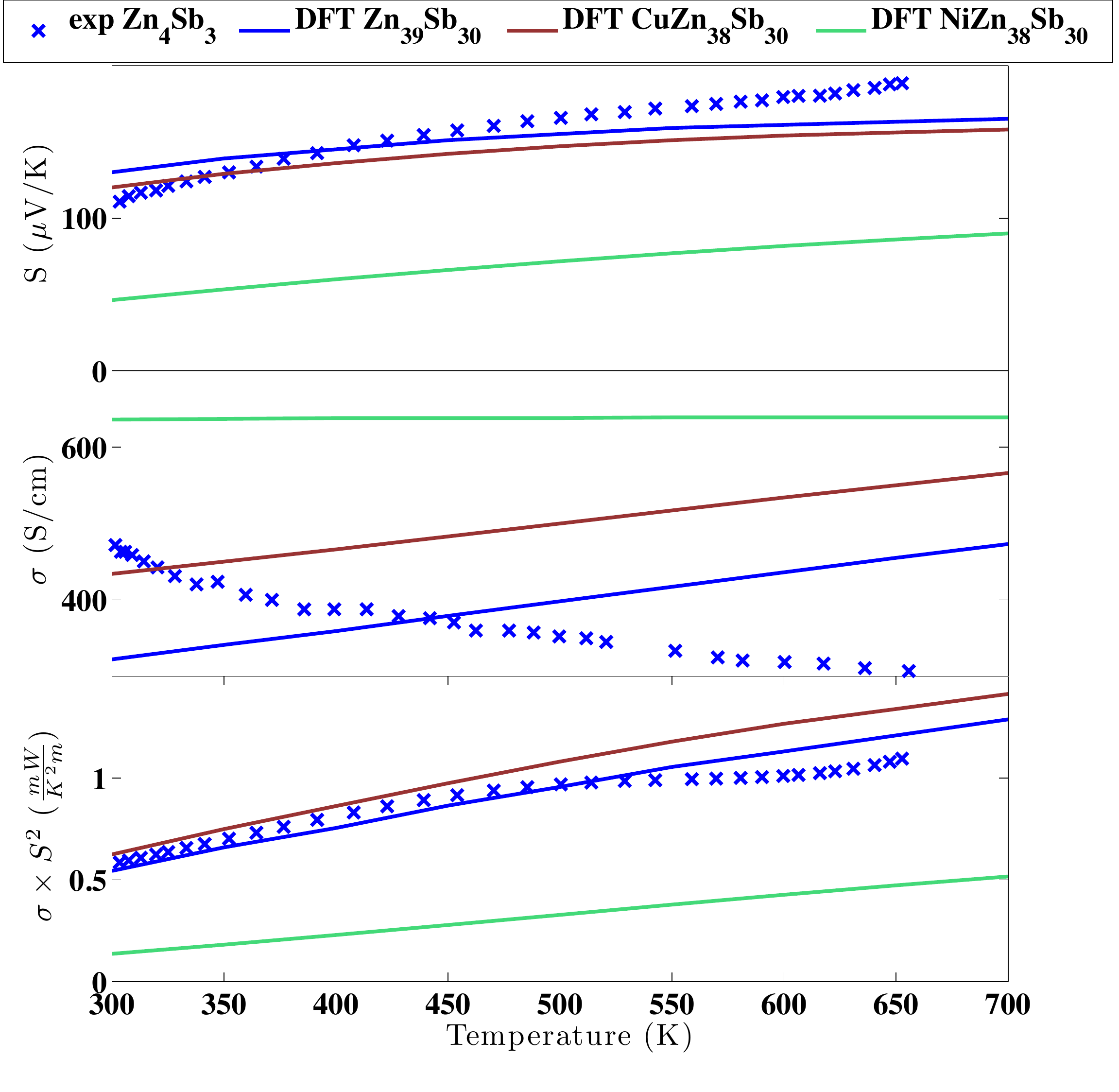}
   \end{center}
   \caption{ \label{fig:Curesults}2.56\% Cu- and Ni-doped Zn$_4$Sb$_3$ calculated and experimental\cite{71} thermoelectric properties: Seebeck coefficient, electrical conductivity, and power factor respectively.}
\end{figure}

We then use the fitting algorithm to rapidly screen various transition metals -- including Co, Fe, Cu, Nb, Ni -- as possible Zn substitutional dopants.  The band structures are shown in Figure \ref{fig:bs1}.  All dopants, except for Cu, induce significant changes in the band structure relative to the undoped compound.  The doped band structures have flatter valence bands, and the presence of midgap states decreases the Seebeck coefficient by an order of magnitude.  Ni and Cu were selected for further study due to the improvement in calculated properties relative to the undoped compound, and the results are seen in Figure \ref{fig:Curesults}.

\begin{figure}
        \subfigure[]{
            \label{fig:dos_undoped}
            \includegraphics[width=0.45\textwidth]{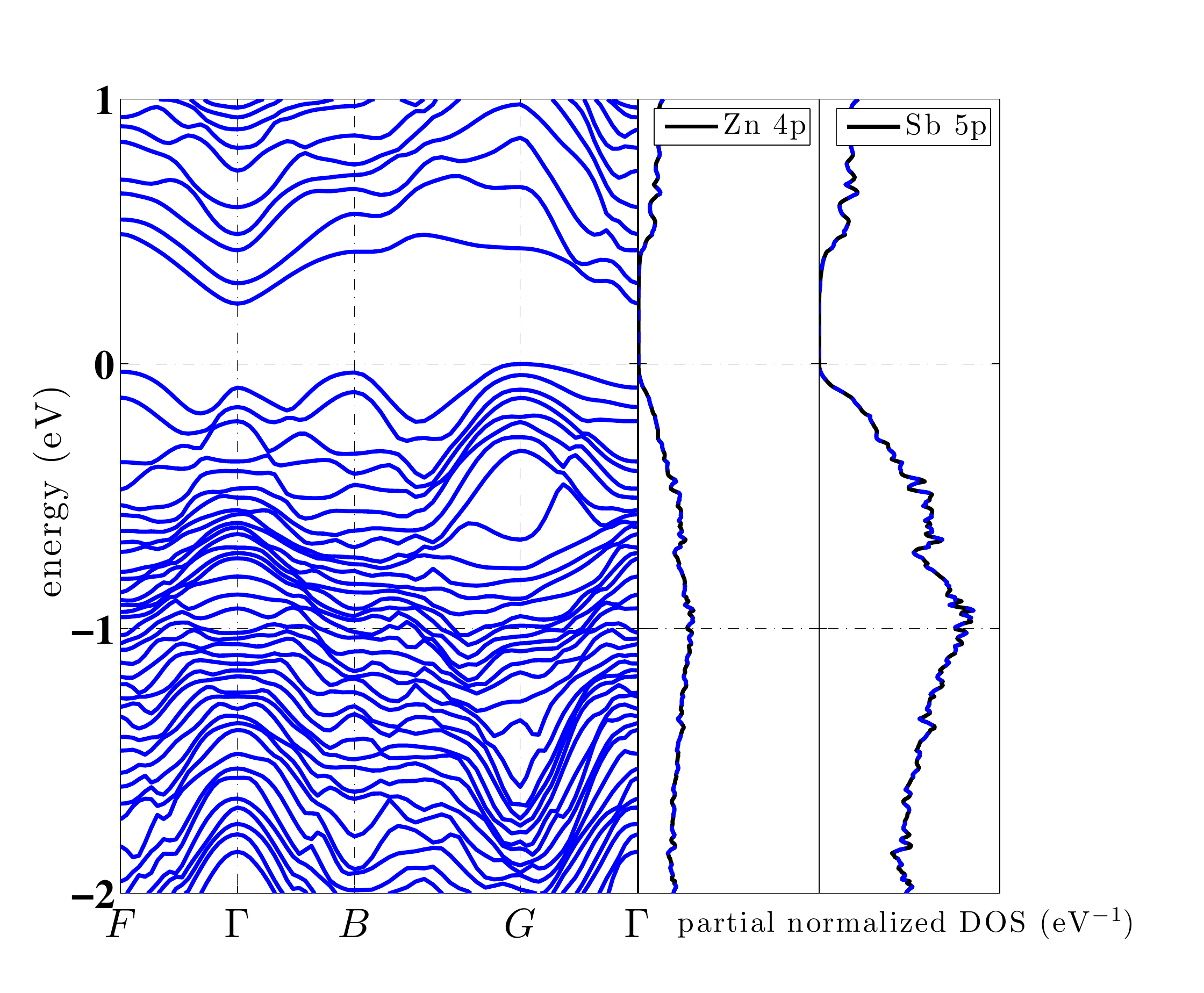}}
       \\ 
        \subfigure[]{
            \label{fig:dos_Cu}
            \includegraphics[width=0.45\textwidth]{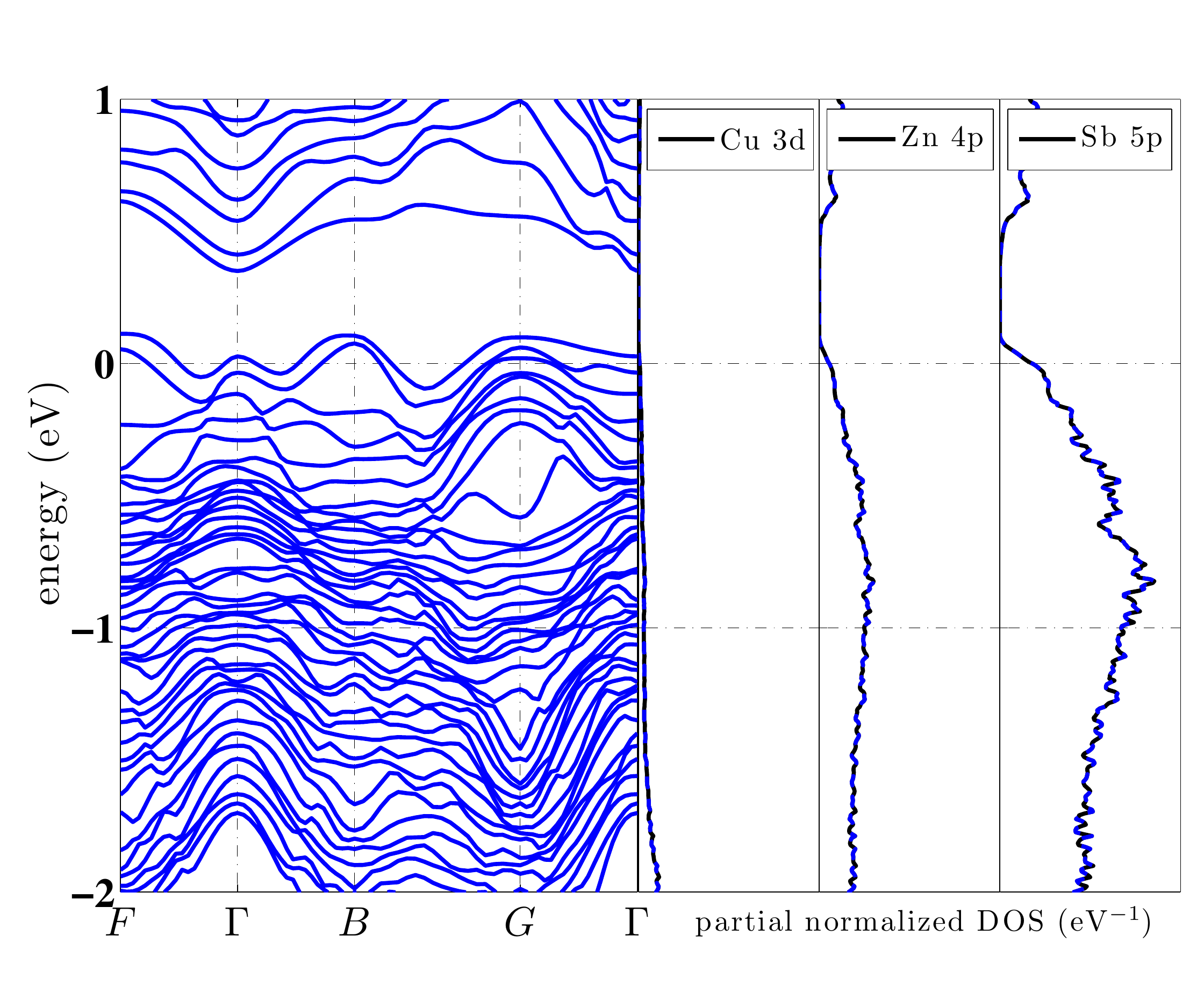}}
	\\
        \subfigure[]{
            \label{fig:dos_Ni}
            \includegraphics[width=0.45\textwidth]{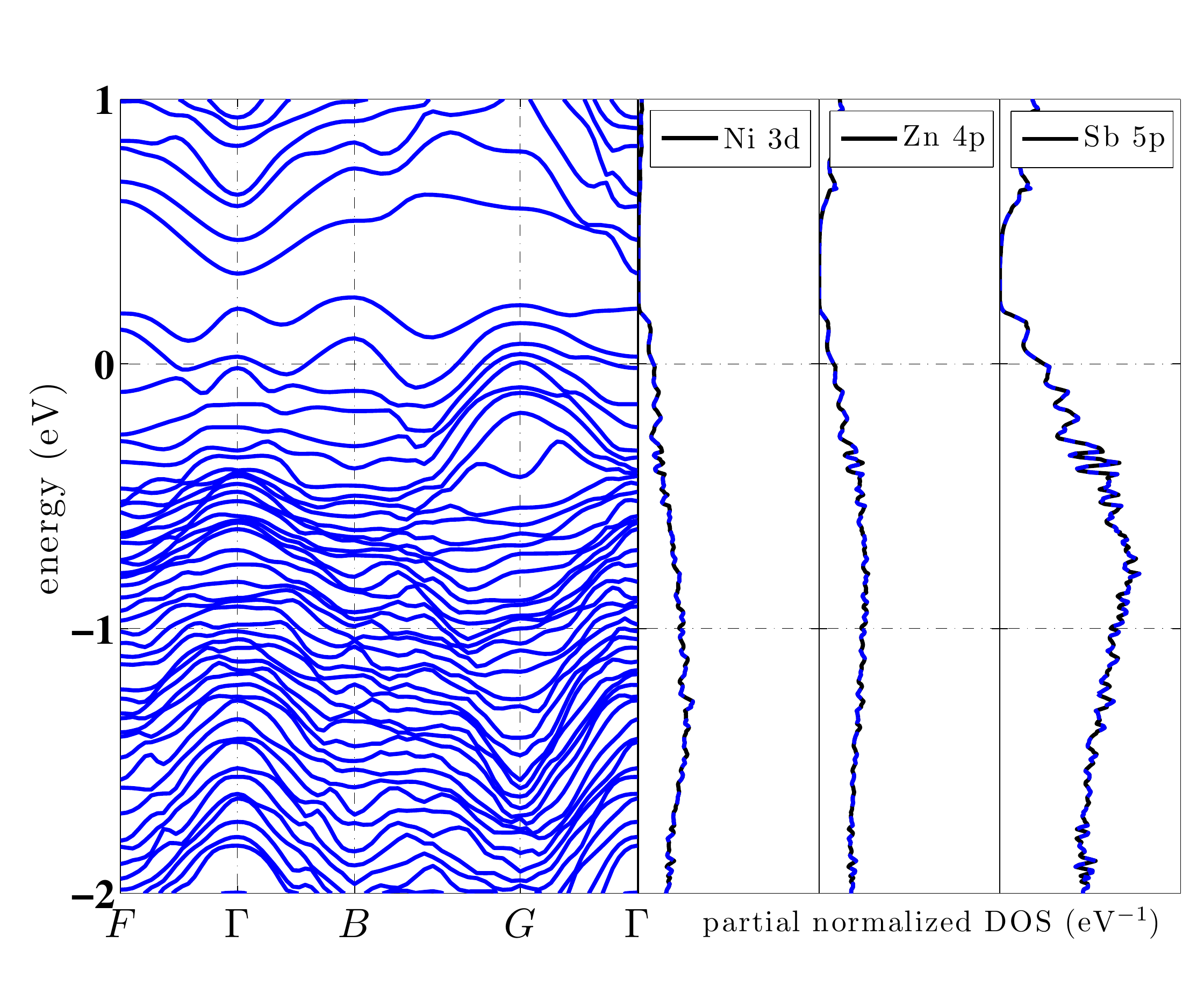}}
    \caption{\label{fig:dos}Band structure and density of states calculated for \subref{fig:dos_undoped} Zn$_{39}$Sb$_{30}$, \subref{fig:dos_Cu} CuZn$_{38}$Sb$_{30}$, and \subref{fig:dos_Ni} NiZn$_{38}$Sb$_{30}$. The number of states are normalized with respect to the total number of electrons in each supercell.}
\end{figure}

The electrical conductivity is improved upon Ni or Cu doping, but the Seebeck coefficient is decreased upon Ni doping.  The appearance of the mid gap state (Figure \ref{fig:dos_Ni}), which decreases the band gap, leads to this decrease in $S$.  Ni states in the VBM significantly increase the electrical conductivity of the material.  On the flip side, Cu does not change the band structure compared to the undoped compound.  Only the core electrons in Cu $d$-orbital are affected upon doping, so no discernible changes in the band structure are noted.  However, the electrical conductivity is increased due to a change in the carrier concentration, as calculated:
\begin{equation}
\label{holeconcentration}
p=\frac{1}{V} \sum_{\epsilon=-\infty}^{\epsilon_v} g_v \left(\epsilon \right)\left[1-f\left(\epsilon \right) \right]\Delta \epsilon
\end{equation}
where $p$ is the hole concentration, $V$ is the unitcell volume, $\epsilon_v$ is the maximum valence energy, $g_v$ is the DFT-calculated density of states for the valence band, and $f(\epsilon)$ is the Fermi-Dirac distribution function. For undoped, Cu-doped, and Ni-doped supercells, $p = 2.41 \times 10^{20} \ \text{cm}^{-3}$ (undoped), $2.93 \times 10^{20} \ \text{cm}^{-3}$ (Cu-doped), and $8.54 \times 10^{20} \ \text{cm}^{-3}$ (Ni-doped). The experimentally \cite{71} measured hole concentration for undoped $\beta-$Zn$_4$Sb$_3$ have been reported as $p = 9.0 \times 10^{19}cm^{-3}$. While the calculated value is slightly different than the experimental value, the increase in carrier concentration by doping has been successfully predicted for doped compounds. Also, as expected, the calculated increase in the carrier concentration is larger for Ni-doped compound than in Cu-doped compound. This is consistent with the changes in the calculated electrical conductivity. We note that we calculated the formation energy of Zn interstitial and Sb vacancy and both are thermodynamically unfavorable thus not decreasing the hole concentration.

\begin{figure}
\begin{center}         
            \includegraphics[width=0.45\textwidth]{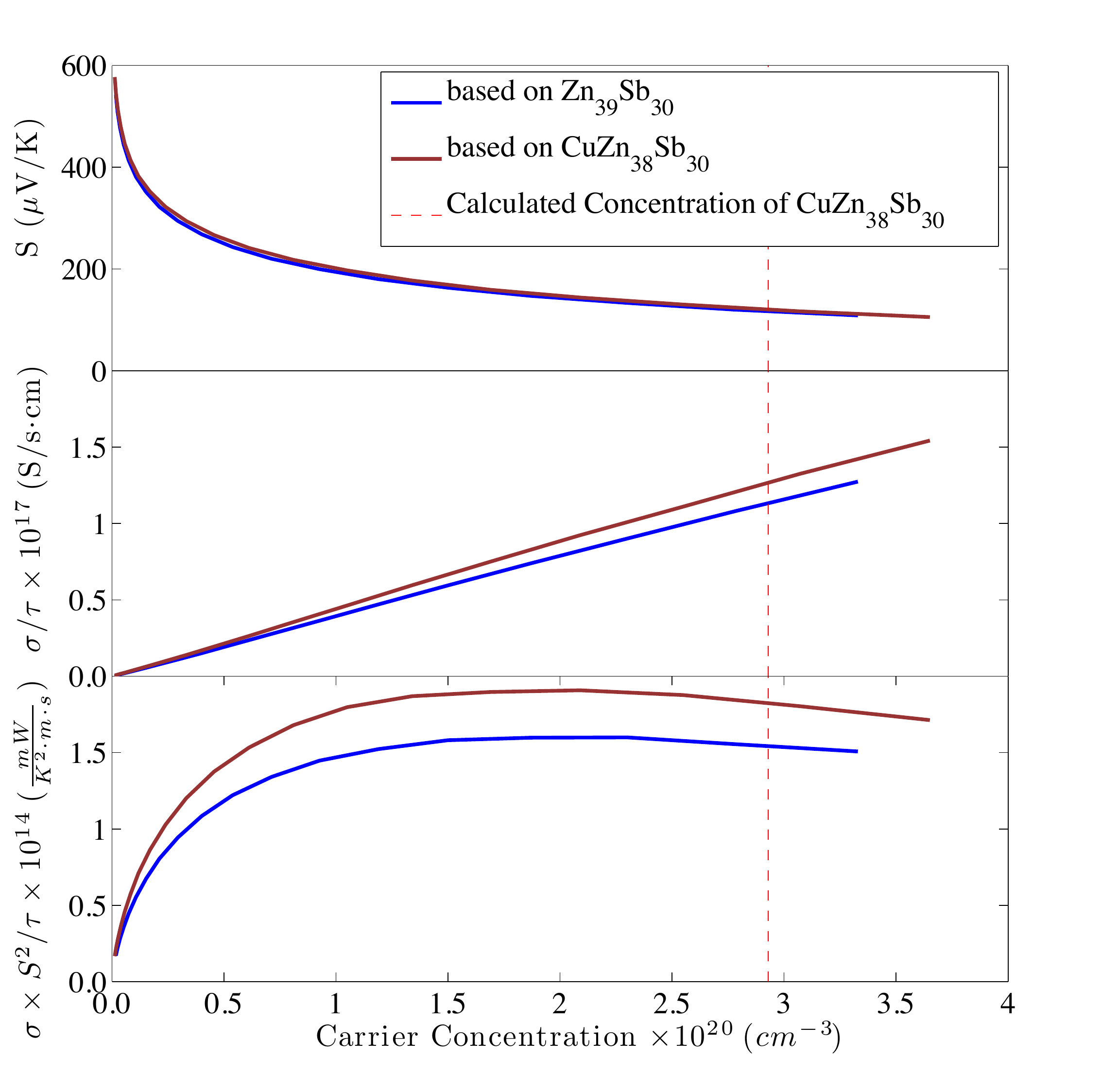}
   \end{center}
   \caption{\label{fig:rigid_band_comparison}Comparison between doping analysis approach presented in this paper and rigid band approximation at 300 K. A difference of 7.2\% was calculated at hole concentration of $2.93\times 10^{20} \ \text{cm}^{-3}$(dashed line) calculated directly for CuZn$_{38}$Sb$_{30}$.}
\end{figure}

As opposed to the usage of the rigid band approximation for doping analysis, in our work, the changes in the band structure resulting from doping are captured by DFT calculations. As shown in Figure \ref{fig:rigid_band_comparison}, this results in a difference of 7.2\% in the calculated power factor divided by relaxation time, between the two methods. Furthermore, it should be noted that within the rigid band approximation, the doping elements and their percentages that create certain carrier concentration cannot be accurately determined, while in the current method, they can be directly calculated via DFT as shown by the dashed line in Figure \ref{fig:rigid_band_comparison}.

\section{Conclusions}
A new method for calculating the properties of lightly doped thermoelectric material using BoltzTraP code was proposed in detail, and validated with Zn$_{39}$BiSb$_{29}$ as an example. We adjust the computed Fermi energy so that the resulting Seebeck coefficient trends with temperature match experimental trends for the undoped compound, and then use this adjusted Fermi energy to compute the electronic relaxation time for both undoped and doped compounds. After screening different elements as possible cation dopants in Zn$_{39}$Sb$_{30}$, this method was used to calculate Seebeck coefficient, electrical conductivity, and power factor of 2.56\% copper and nickel doped zinc antimonide. Copper doped calculation results were more promising compare to nickel. Although copper doping results have been reported previously in simple ZnSb phase, making Cu doped $\beta-$Zn$_4$Sb$_3$ experimentally in different ratios can be the next step considering the fact that more stability of this material was confirmed with formation energies study done in this work.

\subsection{Acknowledgments}
This work used the Extreme Science and Engineering Discovery Environment (XSEDE), which is supported by National Science Foundation grant number OCI-1053575. The authors thank Maria Stoica for helpful discussions and assistance with the post-processing software.

\clearpage

\bibliography{Faghaninia_PhysRevB_2014}

\clearpage

\end{document}